\documentclass[a4paper]{PoS}
\title{The millisecond pulsar contribution to the rising positron fraction}

\ShortTitle{The millisecond pulsar contribution to the rising positron fraction}

\author{\speaker{Christo Venter},$^{a}$ Andreas Kopp,$^{a}$ Alice K Harding,$^{b}$ Peter L Gonthier,$^{c}$ and Ingo B\"{u}sching$^{a}$\\
\llap{$^a$} Centre for Space Research, North-West University, Potchefstroom Campus, Private Bag X6001, Potchefstroom 2520, South Africa\\
\llap{$^b$} Astrophysics Science Division, NASA Goddard Space Flight Center, Greenbelt, MD 20771, USA\\
\llap{$^c$} Hope College, Department of Physics, Holland MI, USA\\
}

\abstract{Pair cascades from millisecond pulsars (MSPs) may be a primary source of Galactic electrons and positrons that contribute to the increase in positron flux above 10 GeV as observed by \textit{PAMELA} and \textit{AMS$-$02}. The \textit{Fermi} Large Area Telescope (LAT) has increased the number of detected $\gamma$-ray MSPs tremendously. Light curve modelling furthermore favours abundant pair production in MSP magnetospheres, so that models of primary cosmic-ray positrons from pulsars should include the contribution from the larger numbers of MSPs and their potentially higher positron output per source. We model the contribution of Galactic MSPs to the terrestrial cosmic-ray electron~/ positron flux by using a population synthesis code to predict the source properties of present-day MSPs. We simulate pair spectra assuming an offset-dipole magnetic field which boosts pair creation rates. We also consider positrons and electrons that have additionally been accelerated to very high energies in the strong intrabinary shocks in black widow (BW) and redback (RB) binary systems. We transport these particles to Earth by calculating their diffusion and the radiative energy losses they suffer in the Galaxy using a model. Our model particle flux increases for non-zero offsets of the magnetic polar caps. We find that pair cascades from MSP magnetospheres contribute only modestly around a few tens of GeV to the measured fluxes. BW and RB fluxes may reach a few tens of percent of the observed flux up to a few TeV. Future observations should constrain the source properties in this case.}

\FullConference{The 34th International Cosmic Ray Conference\\
		30 July- 6 August, 2015\\
		The Hague, The Netherlands}

\begin{document}

\section{Introduction}
We now have firm evidence that the cosmic-ray positron fraction (PF) $\phi(e^+)/[\phi(e^+) + \phi(e^-)]$, with $\phi$ the flux, is an increasing function of energy above $\sim10$~GeV, based on measurements by \textit{PAMELA} \cite{Adriani09,Adriani13}, \textit{Fermi} Large Area Telescope (LAT; \cite{Ackermann12}), and the \textit{Alpha Magnetic Spectrometer} (\textit{AMS$-$02}; \cite{Aguilar13,Aguilar14,Accardo14}). Recent \textit{AMS$-$02} data extended the PF up to 500~GeV, indicating a levelling off of this fraction with energy, as well as being consistent with isotropy.
If the PF is attributed solely to secondary positrons, produced during inelastic collisions between cosmic-ray nuclei and intergalactic hydrogen, this PF is expected to smoothly decrease with energy within the standard framework of cosmic-ray transport (e.g., \cite{Moskalenko98}). The fact that the measured PF rises with energy may therefore point to nearby sources of primary positrons, either from dark matter annihilation (e.g.,~\cite{Grasso09}), or of astrophysical origin. The latter class of sources may include supernovae (e.g.,\cite{Blasi09}), pulsar wind nebulae (e.g., \cite{Serpico12}), young or mature pulsars (e.g., \cite{Gendelev10,Profumo12}), and millisecond pulsars (MSPs; \cite{Kisaka12}). In this paper, we investigate the cosmic-ray flux contribution of the latter source class. 

MSPs are ancient pulsars that have been spun up to short rotational periods $P$ by accretion from a binary companion \cite{Alpar82}. The majority of MSPs were thought to lie below the pair creation death lines assuming dipole magnetic fields \cite{HMZ02}, i.e., being `pair-starved'~\cite{HUM05}. However, detection of narrow, double-peaked $\gamma$-ray light curves trailing the radio peaks, very similar to those of younger pulsars, indicated the existence of narrow accelerator gaps in the MSP magnetospheres, requiring large numbers of electron-positron pairs to screen the electric field parallel to the magnetic field outside these gaps~\cite{Venter09}. Distortions of the surface magnetic field, either in the form of higher multipoles (e.g., \cite{ZC03}) or offset polar caps (PCs; \cite{Arons96,HM11b}), may increase pair production in MSPs. MSPs furthermore produce electron-positron pairs with much higher energies than young pulsars due to their relatively low magnetic fields, so that the MSP pair spectra extend to several TeV~\cite{HM11b}. Additionally, there has recently been a major increase in number of detected MSPs, many of them being nearby and relatively bright. MSPs are thus very promising potential sources of cosmic rays.

In this paper we consider two MSP populations that may contribute to the terrestrial cosmic-ray flux. The first is a Galactic MSP population for which we obtain source properties via population synthesis modelling \cite{Gonthier15} (Section~\ref{sec:synth}). The second population involves MSPs with binary companions. Shocks may form during collision of the the pulsar wind and the companion wind \cite{HG90,Arons93}, possibly accelerating the pairs escaping from the MSP magnetosphere to even higher energies. We therefore also consider black widow (BW) and redback (RB) systems as sources of cosmic rays (Section~\ref{sec:BW}). The number of BWs and RBs has dramatically increased via \textit{Fermi} observations. We calculate the source spectra originating in the MSP magnetospheres (Section~\ref{sec:pairs}) and reaccelerated in binary shocks (for the BWs and RBs; Section~\ref{sec:BW2}). We next transport these spectra through the Galaxy to Earth (Section~\ref{sec:transport}) to assess the MSP contribution to the terrestrial cosmic-ray spectrum (Section~\ref{sec:results}). Our conclusions follow in Section~\ref{sec:concl}. More details may be found in~\cite{Venter15b}.

\section{Source populations}
\subsection{Galactic synthesis model for the present-day MSP population}
\label{sec:synth}
We follow \cite{Gonthier15} to predict the present-day distribution of MSPs (see also \cite{Story07,Venter15b}). The Galaxy is seeded with MSPs (some aged up to 12~Gyr assuming a constant birth rate of $4.5\times10^{-4}$ MSPs per century), which are then evolved in the Galactic potential from their birth location to the present time. We assume that MSPs are ``born'' on the spin-up line, and their surface magnetic field $B_{\rm s}$ does not decay with time. We assume a power-law distribution for $B_{\rm s}$ and adopt a spin-down power of the form (e.g., \cite{Contopoulos14})
\begin{equation}
L_{\rm sd} \sim \frac{2\,\mu^2\, \Omega^4}{3\,c^3}\left(1+\sin^2\alpha\right),\label{eq:Spit}
\end{equation}
where $\mu$ is the magnetic dipole moment, $\Omega$ is the rotational angular velocity, $c$ is the speed of light, and $\alpha$ is the magnetic inclination angle relative to the pulsar's rotational axis.
We constrain the synthesis model parameters by fitting the model output to data from 12 radio surveys as well as from \textit{Fermi} LAT. This simulation predicts the location as well as $P$ and $\dot{P}$ (time derivative of $P$) of $\sim$50~000 Galactic MSPs, which we use as discrete sources of electrons and positrons.

\subsection{MSPs in binary systems -- the BW / RB component}
\label{sec:BW}
About 80\% of known MSPs are in binary systems. A subset of these, the BWs and RBs, may contain strong intrabinary shocks that can further accelerate the pairs. BWs are close binary systems with orbital periods of hours, containing a rotation-powered MSP and a compact companion having very low mass of $\sim 0.01 - 0.05\, M_{\odot}$. The companion stars in BWs undergo intense heating of their atmospheres by the MSP wind, which drives a stellar wind and rapid mass loss from  the star. A shock will form in the pulsar wind at the pressure balance point of the two winds and particle acceleration may occur in these shocks \cite{HG90,Arons93}. RBs are similar systems, except that the companions have somewhat higher masses of $\sim 0.1 - 0.4\, M_{\odot}$ \cite{Roberts11}. The MSPs in both types of system are typically energetic, with $L_{\rm sd} \sim 10^{34} - 10^{35}\,\rm erg\,s^{-1}$. Recent radio searches of \textit{Fermi} unidentified $\gamma$-ray point sources \cite{Ray12} have discovered many new BWs and RBs, with a total of 26 known systems at the present time. By considering only the 24 publicly announced BWs and RBs here, our predictions will be a lower limit to the cosmic-ray flux contribution by binary MSPs.

\section{Source spectra}
\subsection{Injection spectra of particles from the Galactic MSP population}
\label{sec:pairs}
We calculate the spectra of pairs leaving the MSP magnetosphere using a code that follows the development of a PC electron-positron pair cascade in the magnetosphere \cite{HM11b}. A fraction of the curvature radiation photons emitted by primary particles ejected from the stellar surface undergo magnetic pair attenuation \cite{Erber66}. This produces a first-generation pair spectrum which then radiates synchrotron radiation (SR) photons that produce further generations of pairs. The total cascade multiplicity (average number of pairs spawned by each primary lepton) is a strong function of $P$ and $B_{\rm s}$. The sweepback of magnetic field lines near the light cylinder (where the corotation speed equals the speed of light) as well as asymmetric currents within the neutron star may cause the magnetic PCs to be offset from the dipole axis. We adopt a distorted magnetic field structure~\cite{HM11b} that leads to enhanced local electric fields, boosting pair formation even for pulsars below the usual pair death line. We consider PC offset parameter values $\varepsilon=(0, 0.2,0.6)$ and use a grid in $P$ and $B_{\rm s}$ to calculate the source spectrum for each source in the present-day MSP population (Section~\ref{sec:synth}) via interpolation. 
From our simulations we find that about $\sim1$\% of $L_{\rm sd}$ is tapped to generate the pairs. We neglect any further losses of the pair energy before injection into the interstellar medium, since MSPs are not surrounded by nebulae that can degrade the particle energy before escape. 

\subsection{Injection spectra of particles accelerated in intrabinary shocks of BWs and RBs}
\label{sec:BW2}
Pairs escaping from the pulsar magnetosphere may be further accelerated in the intrabinary shock that originates between the pulsar and companion winds in BW and RB systems. The maximum particle energy will be determined by a balance between the minimum acceleration timescale, set by the particle diffusion (which we assume to be Bohm diffusion), and the SR loss timescale. This yields maximum particle energies in the TeV range \cite{HG90}. We assume that the shock-accelerated spectrum will be an exponentially cut off power law with a spectral index of $-2$. We normalize this spectrum by requiring conservation of mass and energy (or equivalently, current and luminosity). We assume a maximum shock efficiency $\eta_{\rm p,max}$ (conversion efficiency of $L_{\rm sd}$ to particle acceleration) of 10\% and 30\%.

\section{Galactic transport of injected leptons}
\label{sec:transport}
In order to transport the injected particles from the MSPs to Earth, we have to make some assumptions regarding the average Galactic background photon and magnetic field energy densities. We approximate the interstellar radiation field using three blackbody components: optical, infrared (IR), and cosmic microwave background (CMB). We use two sets of energy densities, associated with the Galactic Disc and the Galactic Halo \cite{Blies12}. For the Disc, we assume $U_{\rm opt} = U_{\rm IR} = 0.4$~eV\,cm$^{-3}$ and $U_{\rm CMB} = 0.23$~eV\,cm$^{-3}$, while for the Halo we use $U_{\rm opt} = 0.8$~eV\,cm$^{-3}$, $U_{\rm IR} = 0.05$~eV\,cm$^{-3}$, and $U_{\rm CMB} = 0.23$~eV\,cm$^{-3}$.
For the average Galactic magnetic field strength that determines the SR loss rate we use values of $B_{\rm SR}\sim1-3\,\mu$G (\cite{D10}; hereafter D10).
We use a Fokker-Planck-type equation that includes spatial diffusion and energy losses:
  \begin{equation}
\frac{\partial n_{\rm e}}{\partial t}=\mathbf{\nabla}\cdot\left({\cal K}\cdot\mathbf{\nabla} n_{\rm e} \right)-\frac{\partial}{\partial E}\left(\dot E_{\rm total} n_{\rm e}\right)+S,\label{eq:transport}
  \end{equation}
with $n_{\rm e}$ the lepton density (per energy interval). Furthermore, ${\cal K}$ denotes the diffusion tensor and $\dot E_{\rm total}$ the total energy losses (we use SR and inverse Compton (IC) losses, the latter involving the full Klein-Nishina cross section), while $S$ is the source term. Since MSPs are quite old (ages of $\sim10^{10}$~yr), and have very small $\dot{P}$ values, we assume a steady-state scenario and invoke spherical symmetry.
We assume that the diffusion coefficient is spatially independent so that ${\cal K}$ becomes a scalar function of energy only:
  \begin{equation}
  \kappa(E) = \kappa_0\left(\frac{E}{E_{\rm norm}}\right)^{\alpha_{\rm D}}.\label{eq:alpha_D}
  \end{equation}
We use typical values of $\alpha_{\rm D} = 0.3$ or 0.6, $E_{\rm norm} = 1$~GeV, and $\kappa_0 = 0.1~{\rm kpc}^2{\rm Myr}^{-1} \approx 3\times10^{28}$~cm$^2$s$^{-1}$ (e.g., \cite{Moskalenko98}). 
For $S$, we consider $N\sim5\times10^4$ Galactic MSPs from the population synthesis code (Section~\ref{sec:synth}), and $N=24$ for the BW / RB case (Section~{\ref{sec:BW}}). For the $i^{\rm th}$ pulsar in our synthesis population, we assign a pair spectrum $Q_i(P,B_{\rm s},\varepsilon,E)$, as calculated in Section~\ref{sec:pairs} for the corresponding simulated values of $P$, $B_{\rm s}$, and $\varepsilon$. We model this as
\begin{equation}
    S = \sum_i^NQ_i(P,B_{\rm s},E)\delta(\mathbf{r} - \mathbf{r}_{0,i}).
  \end{equation}
Here, $\mathbf{r}_{0,i}$ are the source positions. For a system of infinite extent, Equation~(\ref{eq:transport}) is solved by the following Green's function (e.g., \cite{D10}):
\begin{equation}
G(\mathbf{r},\mathbf{r}_0,E,E_0) = \frac{\Theta(E_0 - E)}{\dot{E}_{\rm total}\left(\pi\lambda\right)^{3/2}}\exp\left(-\frac{|\mathbf{r} - \mathbf{r}_0|^2}{\lambda}\right), \label{eq:Greens}
\end{equation}
with $E_0$ the particle energy at the source, and the square of the propagation scale is given by 
\begin{equation}
\lambda(E,E_0) \equiv 4\int_{E}^{E_0}\frac{\kappa(E^\prime)}{\dot E_{\rm total}(E^\prime)}\,dE^\prime,\\
\end{equation}
and $\Theta(E_0 - E) $ the Heaviside function which ensures that $\lambda>0$.
The lepton flux is given by
\begin{equation}
\phi_e(\mathbf{r},E) = \frac{c}{4\pi}\int\!\!\!\!\int\!\!\!\!\int\!\!\!\!\int G(\mathbf{r},\mathbf{r}_0,E,E_0)S\,dE_0d^3r_0.\label{eq:int}
\end{equation}

\begin{figure}\centering
\includegraphics[width=.7\textwidth]{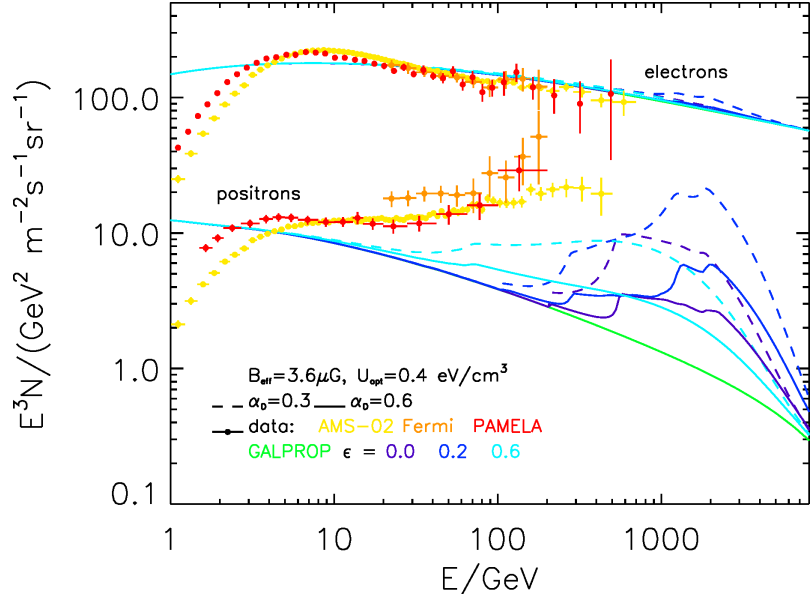}
\caption{Total MSP contribution to the leptonic cosmic-ray spectrum at Earth, for $\kappa_0 = 0.1$~kpc$^2$\,Myr$^{-1}$, $\eta_{\rm p,max}=0.1$, $B_{\rm eff} = 3.6\,\mu$G, and $U_{\rm opt}=0.4$~eV\,cm$^{-3}$. Electron spectra appear at the top and positron spectra at the bottom. The contribution from the synthesis component to the positron spectrum is (marginally) visible at $\sim30$~GeV (for $\varepsilon=0.6$), and that of the BWs and RBs at $\sim1$~TeV. Dashed and solid lines indicate curves for $\alpha_{\rm D}=0.3$ and $\alpha_{\rm D}=0.6$, respectively. The cool colours (purple, blue, and cyan) indicate spectra for $\varepsilon = 0.0, 0.2,$ and $0.6$. Green indicates the ``background'' (non-MSP) electrons and positrons predicted by GALPROP.}
\label{fig14}
\end{figure}

\section{Results}
\label{sec:results}
\begin{figure}\centering
\includegraphics[width=.7\textwidth]{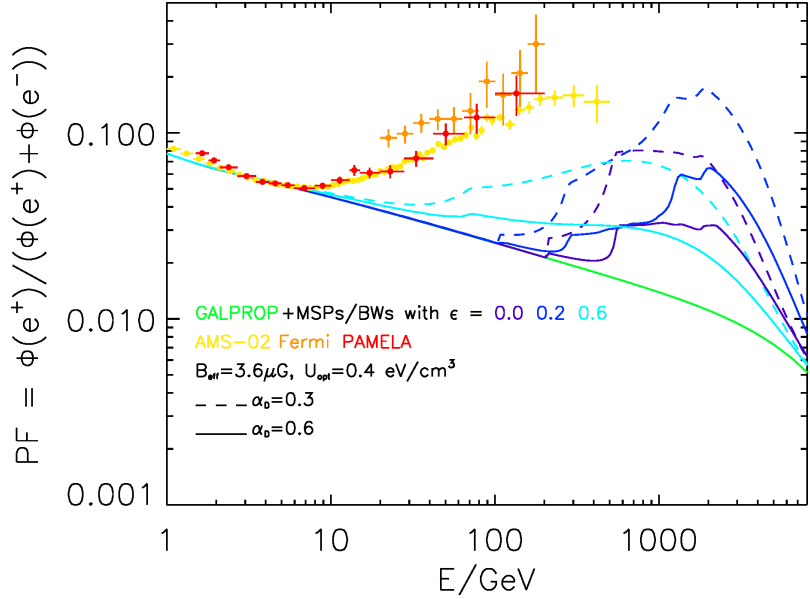}
\caption{Measured \cite{Adriani13,Ackermann12,Accardo14} and predicted PF (including ``background'' contributions from GALPROP in green and the synthesis plus BW / RB contributions in purple, blue, and cyan, indicating $\varepsilon = (0.0, 0.2,0.6$). Here, $\kappa_0 = 0.1$~kpc$^2$\,Myr$^{-1}$, $\eta_{\rm p,max} = 0.1$, $B_{\rm eff} = 3.6\,\mu$G, and $U_{\rm opt}=0.4$~eV\,cm$^{-3}$. Dashed and solid lines are for $\alpha_{\rm D}=0.3$ and $\alpha_{\rm D}=0.6$.}
\label{fig20}
\end{figure}

Figure~\ref{fig14} indicates the ``background'' secondary electron and positron fluxes predicted by GALPROP\footnote{http://galprop.stanford.edu/webrun/} \cite{Vladimirov11} for standard parameters, as well as data from \textit{PAMELA}~\cite{Adriani13}, \textit{Fermi}~\cite{Ackermann12}, and \textit{AMS$-$02}~\cite{Aguilar14}. We show our MSP synthesis spectral contribution (maximal but negligible contribution around 30~GeV) plus BW~/ RB spectral contribution (we assume equal numbers of positrons and electrons) for dipole offsets of $\varepsilon = (0.0, 0.2, 0.6)$,  $\alpha_{\rm D} = (0.3, 0.6)$, and $(B_{\rm eff},U_{\rm opt}) = (3.6\,\mu$G,\,0.4 eV\,cm$^{-3}$). Here, $B_{\rm eff}$ includes the SR and Thomson-limit (non-optical) IC losses which are $\propto E^2 $. We set $\kappa_0 = 0.1$~kpc$^2$\,Myr$^{-1}$ and $\eta_{\rm p,max}=0.1$. The BW / RB contribution becomes higher for a larger $\eta_{\rm p,max}$, and all components are higher for lower values of $\kappa_0$ or $\alpha_{\rm D}$ (due to particle pile-up). The shape of the background model can strongly influence the total lepton spectrum.
Figure~\ref{fig20} shows the measured PF (e.g., \cite{Accardo14}) as well as the GALPROP and synthesis plus BW / RB contributions, for $\kappa_0 = 0.1$~kpc$^2$\,Myr$^{-1}$ and $\eta_{\rm p,max} = 0.1$. The largest contribution is found $\sim100$~GeV when $\varepsilon=0.6$ and $B=3.6\,\mu$G. The highest PF (above 1~TeV) occurs for the lowest values of $\kappa_0$ and $\alpha_{\rm D}$ and highest $\eta_{\rm p,max}$. The BW / RB component makes a significant contribution at a few hundred GeV. Some parameter combinations are excluded by the data, e.g., $\kappa_0 = 0.01$~kpc$^2$\,Myr$^{-1}$, $\eta_{\rm p,max}=0.3$, and $\varepsilon = 0.6$, depending on the background model.

\section{Conclusion}
\label{sec:concl}
We carefully assessed the contribution of MSPs to the cosmic-ray lepton spectra at Earth using a population synthesis code and a pair cascade code to calculate realistic source spectra. We also considered the contribution of binary BW / RB systems, which may further accelerate pairs escaping from the MSP magnetospheres in intrabinary shocks. The predicted MSP particle flux increases for non-zero magnetic field offset parameters $\varepsilon$. This is expected, since larger $\varepsilon$ leads to an increase in the acceleration potential for some regions in magnetic azimuthal phase, implying an enhancement in both the number of particles as well as their maximum energy. The MSPs from the synthesis model make only a modest contribution to the terrestrial cosmic-ray flux at a few tens of GeV, after which this spectral component cuts off. Increased magnetic and soft photon energy densities lead to increased particle energy losses in the Galaxy, and vice versa. The PF is somewhat enhanced above $\sim10$~GeV by this component. The BW / RB component, however, contributes more substantially above several hundred GeV. For some parameter combinations, this component may even exceed the measured positron spectrum, and may violate the PF at high energies, depending on the background model. Alternative sources of primary positrons such as young, nearby pulsars or supernova remnants should also contribute to the cosmic-ray electron and positron flux. Future observations and modeling should continue to constrain the properties of these source classes, as well as improve our understanding of Galactic structure and particles within our Galaxy.

\end{document}